\documentclass[aps,prl,showpacs,floatfix,twocolumn,amsmath,amssymb,superscriptaddress,nobalancelastpage]{revtex4-1}
\usepackage{graphicx}

\begin{document}
\title{Roton-Phonon Interactions in Superfluid $^4$He}
\author{B. F\aa k}
  \affiliation{SPSMS, UMR-E CEA / UJF-Grenoble 1, INAC, F-38054 Grenoble, France}
\author{T. Keller}
  \affiliation{Max-Planck-Institute for Solid State Research, Heisenbergstra{\ss}e 1, D-70569 Stuttgart, Germany}
  \affiliation{FRM-II, TU M\"unchen, Lichtenbergstra{\ss}e 1, D-85747 Garching, Germany}
\author{M. E. Zhitomirsky}
  \affiliation{SPSMS, UMR-E CEA / UJF-Grenoble 1, INAC, F-38054 Grenoble, France}
\author{A. L. Chernyshev}
  \affiliation{Department of Physics and Astronomy, University of California, Irvine, California 92697, USA}
\date{\today}

\begin{abstract}
High-resolution neutron resonance spin-echo measurements of superfluid $^4$He
show that the roton energy does not have the same temperature dependence as the inverse lifetime.
Diagrammatic analysis attributes this to the interaction of rotons with
thermally excited phonons via both four- and three-particle processes,
the latter being allowed by the broken gauge symmetry of the Bose condensate.
The distinct temperature dependence of the roton energy at low temperatures suggests
that the net roton-phonon interaction is repulsive.
\end{abstract}

\pacs{
67.25.dt, 
78.70.Nx, 
03.75.Kk  
}
\maketitle

\begin{figure}[b]
\centering
\includegraphics[width=0.98\columnwidth]{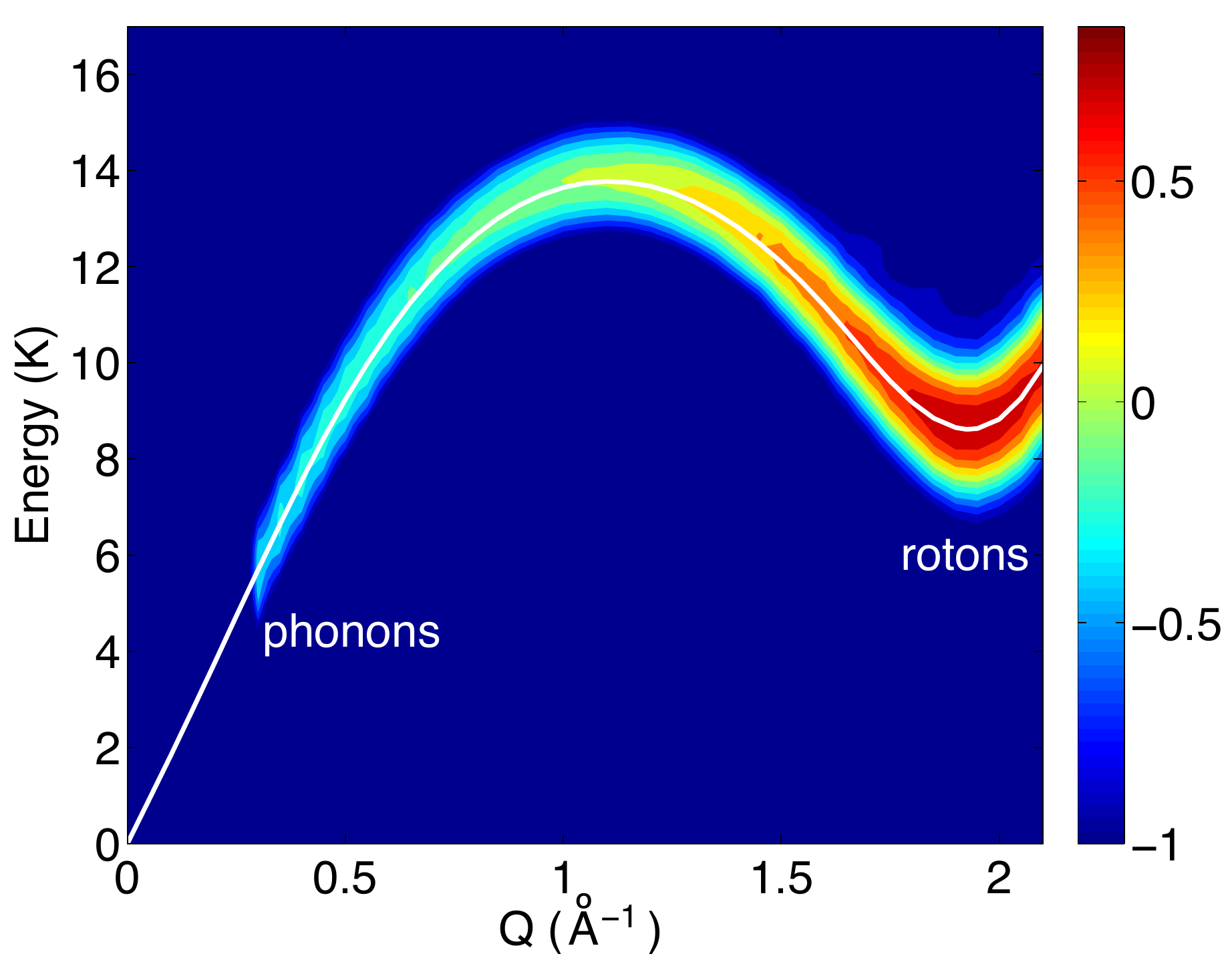}
\caption{The neutron scattering intensity in superfluid $^4$He
at saturated vapor pressure and $T$=1.3 K on a logarithmic scale
as a function of wave vector $Q$ and energy, from Ref.\ \cite{Andersen94a}.
The line shows the dispersion of the quasiparticles,
corresponding to phonons at small wave vectors $Q$
and rotons near the local minimum at $Q_R\approx 1.93$~\AA$^{-1}$\ and energy $\Delta/k_B\approx 8.6$~K.}
\label{FigDisp}
\end{figure}

Superfluid $^4$He is the archetype of a strongly interacting Bose-condensed system.
Its low-temperature properties can be understood in terms of weakly interacting quasiparticles
called phonons and rotons. At zero temperature,  scattering of phonons and rotons
is severely restricted by kinematical conditions due to particularities of their dispersion
depicted in Fig.~\ref{FigDisp} \cite{Andersen94a}, a feature which is intimately related
to superfluid flow. At finite temperatures, rotons acquire a finite lifetime due
to scattering from thermally populated rotons by four-particle processes (4PP),
as predicted in a pioneering work by Landau and Khalatnikov (LK) \cite{LK}
and subsequently observed in neutron scattering experiments \cite{Mezei}.
Surprisingly, manifestations of three-particle processes (3PP)
have not been reported for roton dynamics. Such particle-nonconserving
interactions appear due to a macroscopic occupation of the lowest energy level
and is a hallmark of Bose-Einstein condensation \cite{Volovik03}.

In this Letter, we combine high-resolution neutron scattering measurements with
theoretical  analysis to show that roton-phonon processes in superfluid $^4$He give
rise to a distinct temperature dependence of the roton energy, $\Delta(T)$,
at sufficiently low temperatures.
We use the neutron resonance spin-echo method  \cite{Golub87,Keller98},
which due to its extremely stable spin-echo phase
allows to measure the energy {\it shift} of an excitation with high precision
simultaneously with the linewidth.
The improved resolution in our setup
makes it possible to study  effects
that are only revealed  at  temperatures below 1~K,
where the roton-roton scattering due to 4PPs predicted by LK
no longer dominates other  processes.
The main result of our work is that the roton linewidth and energy shift in superfluid $^4$He
have {\it different} temperature dependencies, in particular at low temperatures.
This is in contrast to the LK theory \cite{LK} and subsequent theoretical work \cite{BPZ},
where the roton lifetime and energy shift are predicted to have the same temperature dependence,
both simply proportional to the thermal roton population.
A systematic evaluation of the leading self-energy diagrams show that the
experimental results can be understood in terms of the interaction of rotons
with thermally excited phonons via both three- and four-particle processes.
This development opens up new perspectives for studies of quasiparticle
interactions in a variety of Bose-condensed systems.

The measurements were performed on the TRISP
thermal polarized neutron triple-axis spectrometer at the FRM-II reactor (Munich),
where radio-frequency flippers are used to encode the energy of each neutron
in its Larmor precession angle, thereby improving the energy resolution by two orders of magnitude.
Our setup allows to measure energy shifts with an unprecedented precision of
2 mK ($\sim\!0.2$ $\mu$eV)
and linewidths down to 1.5 mK ($\sim\!0.17$ $\mu$eV).
The polarization and energy of the incoming neutrons
were determined by a supermirror guide
and a PG(002) monochromator, respectively,
with the higher-order neutrons eliminated by a velocity selector.
Analysis of the final energy (6 meV) and polarization of the scattered neutrons
were done with a PG(002) crystal assembly and
a supermirror bender, respectively.
The radio-frequency flippers were inserted in zero-field devices
mounted before and after the sample.
TRISP was operated in both boot-strap and non-boot-strap mode \cite{Gahler88}
in  \verb/--+/ configuration
and radio frequencies between 60 and 292.62 kHz,
which cover spin-echo times 15$\leq\!\tau\!\leq$190 ps.
The superfluid $^4$He sample, held in an aluminium cylinder
of inner diameter 26 mm and volume 30 cm$^3$
equipped with Cd spacers to reduce multiple scattering,
was cooled to below 0.5 K using a $^3$He insert in a pulse-tube cryocooler.
Typical count rates for the roton was 60 counts per minute.

\begin{figure}
\centering
\includegraphics[width=0.98\columnwidth]{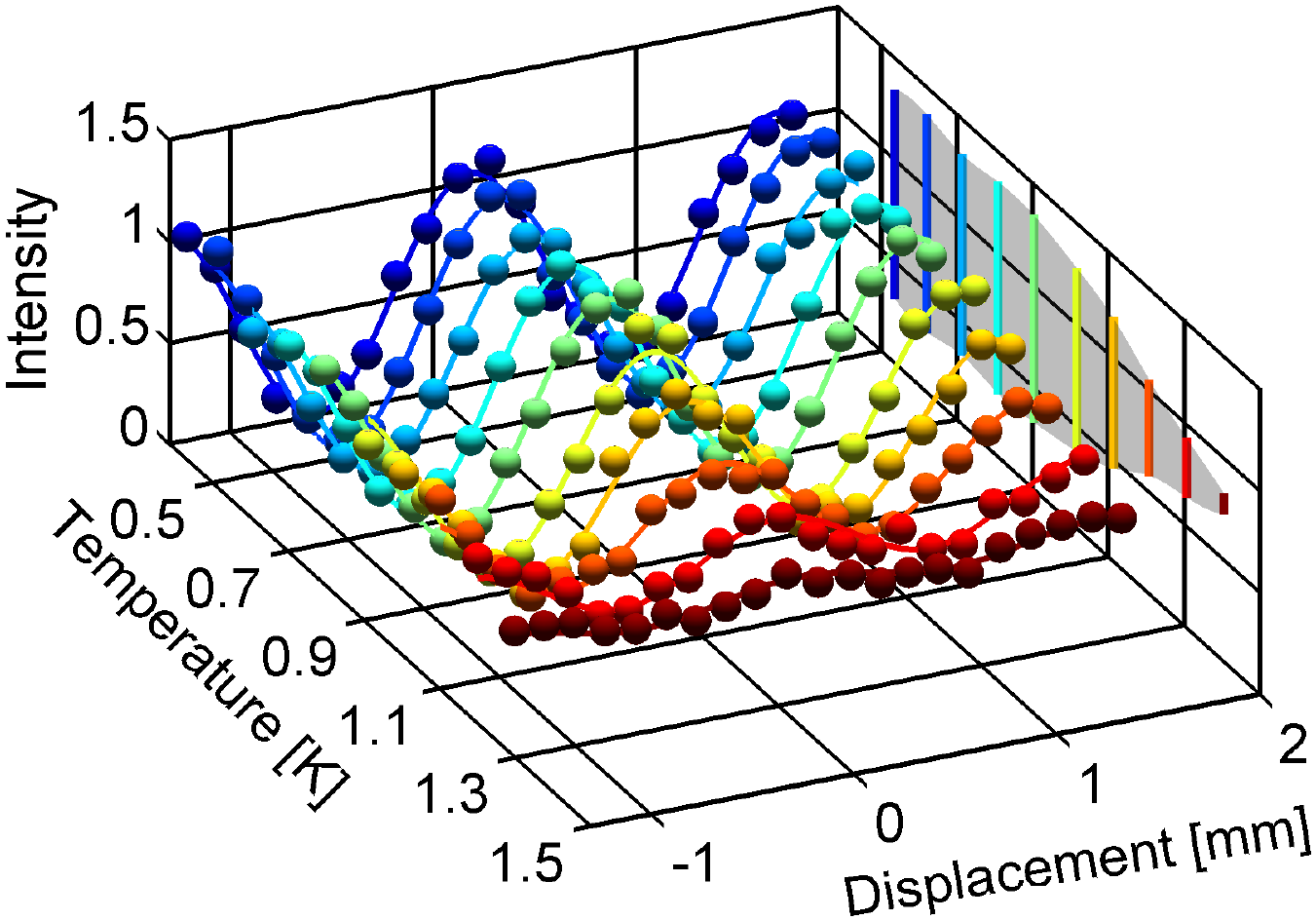}
\caption{Spin-echo scans as a function of coil displacement and temperature.
With increasing temperature,
the neutron polarization  decreases
(shown by the projected grey shadow)
and the phase of the oscillation shifts.
The statistical errors in the neutron count rates are smaller than the symbol size.
The lines are fits by a cosine function. }
\label{FigEcho}
\end{figure}

The intensity of the scattered neutrons is recorded
as a function of the length of one of the zero-field precession paths
by scanning the displacement of one resonance coil.
The oscillatory pattern can be fitted by a cosine function (see Fig.~\ref{FigEcho})
where the amplitude of the oscillatory component
is a measure of the neutron polarization
and gives the lifetime (inverse linewidth) of the excitation
while the phase of the oscillations (position of maximum intensity)
gives the relative energy of the excitation.

The roton energy shift 
\begin{equation}
\delta(T) = \Delta(T) - \Delta(0)
\end{equation}
was obtained in this way for temperatures between 0.5 and 1.6 K
using spin-echo times $\tau$ of 50 and 115 ps.
Since our measurements determine changes in the roton energy,
and not the energy in absolute units, we take $\delta(T)$=0
at the lowest temperature of the measurements, $T$=0.5~K.
To measure the roton linewidth $\Gamma(T)$ (half width at half maximum),
we first verified that the decay of the roton is exponential,
by measuring the neutron polarization
$P(\tau,T)\!=\!P_0\exp[-\Gamma(T)\tau]$
for spin-echo times $\tau$ between 16 and 190 ps
at six temperatures between 0.5 and 1.5 K.
Once the exponential decay was established,
further measurements of $P(\tau,T)$ at  intermediate temperatures
were performed for spin-echo times of 50, 115, and 190 ps.
The linewidth obtained from these data sets was put on an absolute scale
using $P_0$ obtained from direct beam calibration measurements
and taking into account the depolarization due to beam divergence
and dispersion curvature using the formalism of Ref.~\cite{Habicht03}.
At $T$=0.5~K,
where no measurable broadening of the roton line is expected,
we find a negligible linewidth ($<$2 mK),
which confirms this calibration procedure.

\begin{figure}
\centering
\includegraphics[width=0.98\columnwidth]{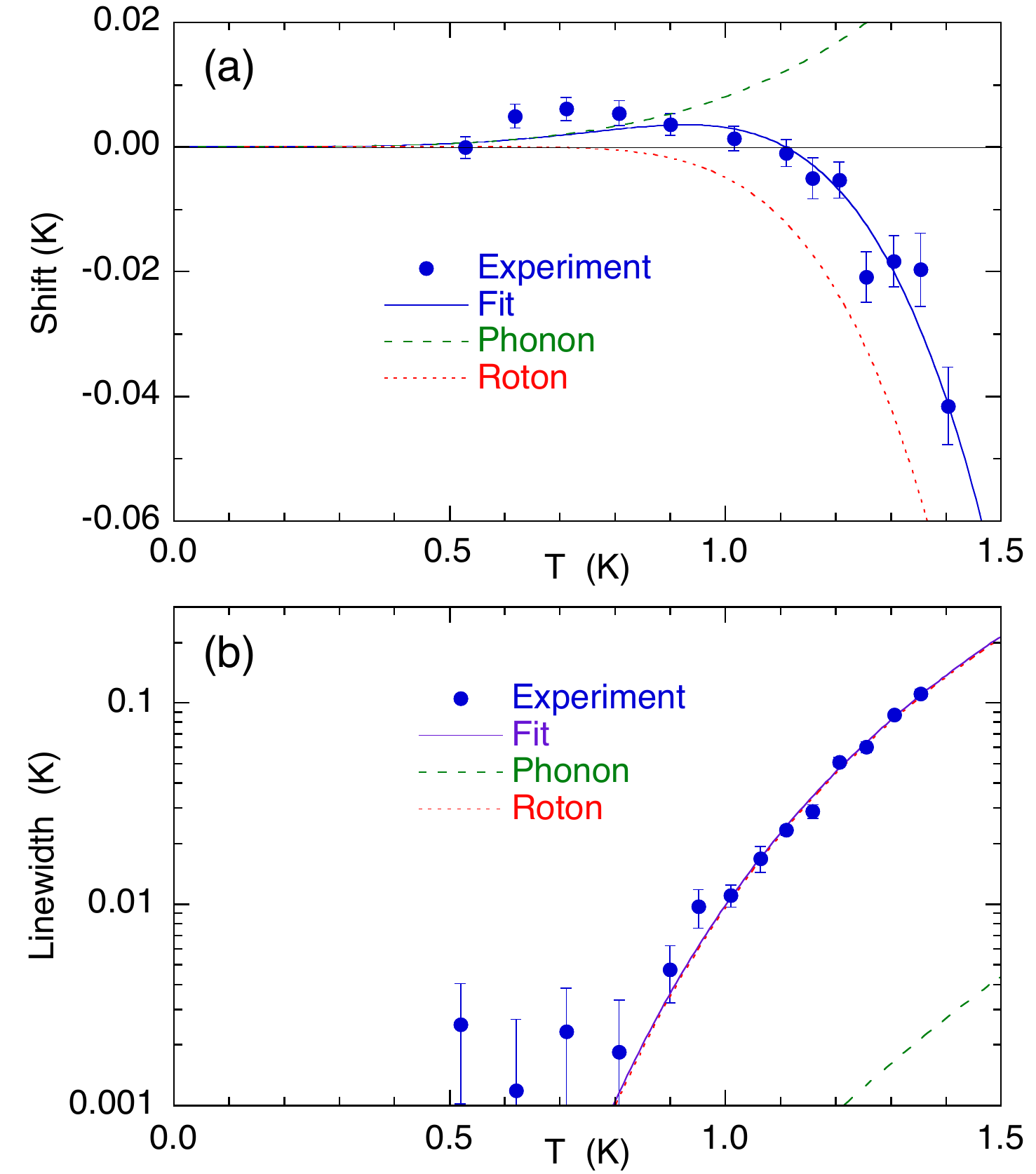}
\caption{Temperature dependence of (a) the roton energy shift
$\delta(T)$ on a linear scale
and (b) of the roton linewidth $\Gamma(T)$
(full width at half maximum) on a logarithmic scale.
Symbols are experimental data
with one-sigma error bars,
solid line the best least-squares fit of Eqs.~(\ref{EqG}) and (\ref{EqD}),
dotted line the roton contribution, and dashed line the phonon contribution.
 }
\label{FigResult}
\end{figure}

The experimental results are shown in Fig.~\ref{FigResult}.
The roton linewidth, Fig.~\ref{FigResult}(b), follows the temperature dependence
given by roton-roton scattering
that has been observed previously \cite{Mezei,Andersen96,Farhi}
and agrees with the LK theory \cite{LK,BPZ}.
On the other hand, the roton energy initially {\it increases} with temperature,
see Fig.~\ref{FigResult}(a),
before it begins a monotonic decrease at $T>1$ K.
The change of sign of the roton energy shift
provides unambiguous evidence that
more than one interaction process contributes to it.
In one earlier report \cite{Andersen96},
the roton energy was found to change more slowly
than predicted by the LK theory \cite{BPZ}.
However, no change of the sign of the energy shift was observed
and no explanation of the behavior was found at that time.

\begin{figure}
\center
\includegraphics[width=0.98\columnwidth]{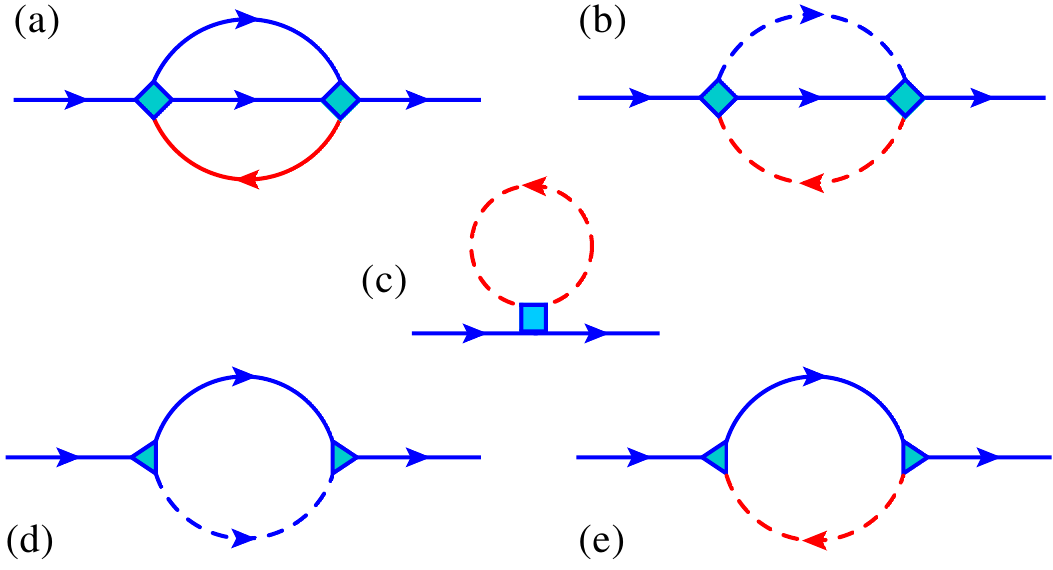}
\caption{Self-energy diagrams due to quasiparticle interactions in superfluid
$^4$He. Solid lines correspond to rotons and dashed lines to phonons. Red lines
indicate thermally excited quasiparticles. Triangles and squares correspond to
cubic and quartic vertices, respectively. Four-particle processes:  (a)
A roton  is scattered by a thermal roton or (b) by a thermal phonon. In (c), the roton
interacts with a uniform density of thermal phonons (the Hartree term).
Three-particle processes: (d) decay and (e) coalescence of a roton
with a thermal phonon.}
\label{FigDiag}
\end{figure}

In order to explain these experimental findings,
we use a diagrammatic approach \cite{Mahan} to evaluate the contributions
from the lowest-order self-energy diagrams shown in Fig.~\ref{FigDiag}.
In contrast to previous work,
we explicitly take into account the interaction of a roton
with thermally excited low-energy {\em phonons},
and include both three- and four-particle processes.

The 4PP interaction in the diagrams of Figs.~\ref{FigDiag}(a)--\ref{FigDiag}(c) is given by
\begin{equation}
\hat{V}_4 = \sum_{ {\bf p}, {\bf q}, {\bf p}', {\bf q}'}
V_{ {\bf p}{\bf q};{\bf p}'{\bf q}'} a^\dagger_{{\bf p}'}a^\dagger_{{\bf q}'}a_{{\bf q}}a_{{\bf p}}
\label{EqV4}
\end{equation}
with $\bf p+q=p'+q'$. The quartic vertex $V_{ {\bf p}{\bf q};{\bf p}'{\bf q}'}$
is a constant for roton-roton scattering and
$V_{ {\bf p}{\bf q};{\bf p}'{\bf q}'}\sim \sqrt{qq'}$ for
roton-phonon interaction processes \cite{LK,BPZ},
where ${\bf q}$ and ${\bf q}'$ are the phonon momenta.
We assume that the energy of quasiparticles
with momentum $p$ close to the roton minimum $p_R$
can be parameterized by the Landau spectrum
$\epsilon_p=\Delta + (p-p_R)^2/(2m^\star)$
with $m^\star$ being the roton mass,
while quasiparticles in the phonon region with momentum $q$
are assumed to have a linear dispersion $\omega_q=cq$.
We use numerical values of
$\Delta/k_B  = 8.622$~K \cite{Andersen94a},
$p_R/\hbar = Q_R  = 1.931$~\AA$^{-1}$  \cite{Andersen94a},
$\mu=m^\star/m_{\rm He}=0.144$ \cite{Andersen94a},
$c = 238.3$~m/s \cite{Abraham70},
and $\rho= 0.14513$~g/cm$^3$  \cite{Kerr64},
and express linewidths and energy shifts in units of Kelvin.

{\em Roton-roton scattering},
i.e., the scattering of a roton by a thermally excited roton,
is depicted as a self-energy diagram in Fig.~\ref{FigDiag}(a).
The roton energy shift and the linewidth due to this process
using Eq.~(\ref{EqV4}) are both proportional to the number of thermally excited rotons,
\begin{equation}
N_r(T) \propto   \sqrt T \left[ 1 + \alpha \sqrt{\mu T}  \right]e^{-\Delta(T)/T}  ,
\label{EqNR}
\end{equation}
as first shown by LK  \cite{LK}.
The minor correction factor  in brackets is due to
small deviations of the  roton dispersion from a parabolic form,
with $\alpha \sqrt\mu=0.0603$ K$^{-1/2}$ \cite{BPZ,BPF}.

{\em Roton-phonon scattering}, i.e.,
the scattering of a roton by a thermally excited phonon,
is represented by the diagram in Fig.~\ref{FigDiag}(b),
which is essentially the same as that for roton-roton scattering.
However, since phonons are gapless,
the linewidth and the energy shift of a roton due to this scattering process
depend on temperature via a power law.
The linewidth using Eq.~(\ref{EqV4}) is
\begin{eqnarray}
\Gamma_{\bf p}(T)&=&  \pi \sum_{\bf q,q'} |
V_{ {\bf p}{\bf q};{\bf p}'{\bf q}'}|^2 \Big[n_{\bf q}(1+ n_{{\bf q}'}+n_{{\bf p}'})
\nonumber\\
&&-n_{{\bf q}'}n_{{\bf p}'}\Big] \, \delta(\epsilon_{\bf p}+
\omega_{\bf q}-\omega_{{\bf q}'} -\epsilon_{{\bf p}'}) ,
\label{EqG4}
\end{eqnarray}
which yields the low-$T$ asymptote
\begin{equation}
\frac{\Gamma_P(T)}{k_B}
\approx \frac{\pi^3}{21} \, \tilde{A}^2 \,
\frac{p_R^2 k_B^6 T^7}{\rho^2 \hbar^6 c^8} =
1.4 \times 10^{-4} \tilde{A}^2 \, T^7,
\label{EqT7}
\end{equation}
where $\tilde{A}$ is a dimensionless constant characterizing the roton-phonon interaction.
It can be evaluated in terms of density fluctuations of the phonon field
in the quantum hydrodynamic limit as \cite{LK}
\begin{equation}
\tilde{A}= \frac{\Delta}{p_R  c}
\left( \frac{\rho^2}{\Delta} \, \frac{\partial^2\Delta}{\partial\rho^2}\right) +
\frac{p_R}{c m^\star}
\left( \frac{\rho}{p_R} \, \frac{\partial p_R}{\partial\rho} \right)^2  .
\label{EqTA}
\end{equation}
The numerical value of $\tilde{A}$ is not known with precision.
Early estimates indicated a negative value $-0.1$ \cite{KC},
while more recent neutron scattering measurements
give values in the range $-0.38$ to $+1.35$ \cite{Dietrich72,Gibbs99}.
The uncertainty comes mostly from the roton mass
and from the density dependence of the roton energy $\Delta(\rho)$.
Using values of $\tilde{A}$ in this range,
the prefactor of the $T^7$-term in Eq.~(\ref{EqT7}) turns out to be so small
that this process is negligible compared to that of the roton-roton scattering [Eq.~(\ref{EqNR})]
down to $T=0.5$~K,
where both mechanisms yield an undetectable linewidth of
the order of $10^{-6}$ K.
Qualitatively, the interaction of the massless long-wavelength phonon with
a massive almost immobile roton is analogous to Rayleigh scattering.
Multiplying the corresponding cross-section
$\sigma \propto \omega^4\sim T^4$ by the density of thermally excited
phonons, $N_{ph}(T)\propto T^3$, leads to a $T^7$-law.
The  roton energy shift evaluated from
the roton-phonon diagram in Fig.~\ref{FigDiag}(b) has an equivalent $T^7$ contribution,
which can also be neglected compared to
the $T^4$ contributions from other interaction processes that are discussed next.
We also note that a $T^7$ power law for  the energy shift
is not consistent with our experimental data.

{\em The Hartree term},
a 4PP illustrated in the diagram of Fig.~\ref{FigDiag}(c),
contributes only to the real part of the self-energy.
It corresponds to an energy renormalization of a roton
propagating in a gas of thermally excited phonons.
Straightforward calculation of  this diagram using Eq.~(\ref{EqV4})
gives the energy shift
\begin{equation}
\delta_4(T)=\Sigma_H({\bf p}) \approx \sum_{\bf q}
V_{ {\bf p}{\bf q};{\bf p}{\bf q}}  n_{\bf q}
\label{EqD4}
\end{equation}
with a low-temperature asymptote of the form
\begin{equation}
\frac{\delta_4(T)}{k_B}
\approx \frac{\pi^2}{60} \, \tilde{A} \, \frac{p_Rk_B^3T^4}{\rho \hbar^3 c^4}
= 1.6 \times 10^{-3} \tilde{A} \, T^4.
\label{EqHartree}
\end{equation}

{\em Three-particle processes} are depicted in Figs.~\ref{FigDiag}(d) and \ref{FigDiag}(e),
where the first diagram corresponds to decay of
a roton into a roton and a phonon
and the second diagram describes absorption
of a thermal phonon by a roton.
For the dispersion shown in Fig.~\ref{FigDisp},
these three-particle processes do not simultaneously conserve energy and momentum
and do not contribute to the roton lifetime,
i.e., the resulting roton self-energies have no imaginary part.
However, the real part of the self-energy, which
corresponds to virtual transitions into higher-energy states,
is non-zero and gives rise to
a temperature-dependent energy shift.
The 3PP interaction is given by
\begin{equation}
\hat{V}_3 = \sum_{ {\bf p}, {\bf q}, {\bf p}'}
V_{ {\bf p};{\bf q}{\bf p}'} a^\dagger_{\bf p'}a^\dagger_{\bf q} a_{\bf p}
+ \textrm{h.\,c.}
\label{EqV3}
\end{equation}
with $\bf p=q+p'$.
For interaction processes involving one phonon $\bf q$, the cubic vertex
scales as $V_{ {\bf p};{\bf q}{\bf p}'} \propto \sqrt{q}$
(for the full expression, see Ref.~\cite{LK}).
The net contribution of the 3PP diagrams in Figs.~\ref{FigDiag}(d) and \ref{FigDiag}(e) to
the temperature dependent part of the roton energy is
\begin{equation}
\delta_3(T) = \sum_{\bf q} \big|V_{ {\bf p};{\bf q}{\bf p'}}\big|^2  \left[
\frac{n_{\bf q}+n_{\bf p'}}{\epsilon_{\bf p}\!-\!\omega_{\bf q}\!-\!\epsilon_{\bf p'}}\!+\!
\frac{n_{\bf q}-n_{\bf p'}}{\epsilon_{\bf p}\!+\!\omega_{\bf q}\!-\!\epsilon_{\bf p'}}\right].
\label{EqD3}
\end{equation}
At low-temperatures, Eq.~(\ref{EqD3}) gives a $T^4$ law,
\begin{equation}
\frac{\delta_3(T)}{k_B}
= \frac{\pi^2}{300} \, \frac{p_R^2k_B^3T^4}{\hbar^3 \rho m c^5}
= 3 \times 10^{-3} \, T^4,
\label{EqT4}
\end{equation}
similar to the Hartree term of Eq.~(\ref{EqHartree}) but with a different prefactor.
While the contribution of Eq.~(\ref{EqT4}) to the energy shift is positive,
that of Eq.~(\ref{EqHartree}) depends on the sign of the roton-phonon interaction.

We conclude from these theoretical considerations
that the dominating process contributing to the
temperature-dependent linewidth of the roton,
$\Gamma(T)$, is the roton-roton scattering,
while the shift of the roton energy,
$\delta(T)$,
is due to both roton-roton scattering
and the renormalization of the roton energy
due to the background of thermally excited phonons.
Thus, we obtain
\begin{equation}
\Gamma(T) =  \gamma_P T^7
+ \gamma_R  \sqrt T
\left[ 1\!+\!\alpha \sqrt{\mu T}  \right]e^{-\Delta(T)/T}
\label{EqG}\\
\end{equation}
and
\begin{equation}
\delta(T) = \delta_P T^4
- \delta_R  \sqrt T
\left[ 1+\alpha \sqrt{\mu T}\right]e^{-\Delta(T)/T}.
\label{EqD}
\end{equation}

The four parameters $\gamma_P$, $\gamma_R$, $\delta_P$, and $\delta_R$
were determined by fitting
Eqs.~(\ref{EqG}) and (\ref{EqD}) self-consistently
[since $\Delta(T)$ depends on Eq.~(\ref{EqD})]
to our experimental data.
The lines in Fig.~\ref{FigResult} show the resulting least-squares fit.
Clearly, the  $T^4$-term
explains the upturn of the roton energy
at the lowest temperatures, before the downturn above 1 K,
where roton-roton scattering dominates.
For the roton linewidth, inclusion of the $T^7$ term
due to roton-phonon scattering processes
does not improve the description of the experimental data
in the measured temperature regime, $T>0.5$ K.

The best fit gives roton-roton scattering contributions of
$\gamma_R = 49.6 \pm 1.1$ K$^{1/2}$
and $\delta_R = 25.6 \pm 3.0$ K$^{1/2}$,
in a good agreement with previous experimental determinations,
which are in the ranges
$\gamma_R\approx 42-47$
and $\delta_R\approx 19-25$,
respectively \cite{Mezei,BPZ,Andersen96,Farhi}.
The $T^7$ roton-phonon contribution to the roton linewidth is not significant,
as shown in Fig.~\ref{FigResult}(b),
where we have plotted as a dashed line an estimate of this contribution,
taking the largest  value of
the roton-phonon interaction parameter
compatible with the measured density dependence of the roton parameters,
$\tilde{A} = 1.35$.
On the other hand, the $T^4$-contribution to the roton energy is clearly significant
in the experimental fit, see Fig.~\ref{FigResult}(a),
yielding a value of $\delta_P = 8.0 \pm 1.5 \times 10^{-3}$ K$^{-3}$.
Our theoretical estimate from Eqs.~(\ref{EqHartree}) and (\ref{EqT4}) with
$\tilde{A}=1.35$ gives
 $\delta^{\rm th}_P = 5 \times 10^{-3}$ K$^{-3}$,
close to the experimentally observed value.
Vertex corrections to the diagrams in Figs.~\ref{FigDiag}(d) and \ref{FigDiag}(e),
which are beyond the scope of the present theoretical calculations,
increases the value of $\delta_3$
and would further improve the agreement.

In conclusion, our work shows that at sufficiently low temperatures,
the roton energy is affected not only by scattering from thermally excited rotons
but also by the presence of a
``thermal bath'' of  excited phonons.
The effect is particularly visible since the two processes give rise
to different signs of the energy shift.
The magnitude of the initial upward shift of the roton energy
suggests that it has contributions from both
the three-particle cubic vertex terms of Figs.~\ref{FigDiag}(d) and \ref{FigDiag}(e)
and from the  quartic Hartree-term of Fig.~\ref{FigDiag}(c).
We further conclude that the
roton-phonon interaction parameter $\tilde{A}$ is likely to be positive,
and not negative as was assumed until now.
This implies that the effective roton-phonon interaction is repulsive,
in contrast to the roton-roton interaction, which is attractive \cite{Pistolesi}.
The roton-phonon interaction does not give any observable roton linewidth.
The present results show the importance
of both cubic and quartic vertices in superfluid $^4$He,
and we expect that this will have a wide-ranging impact
for Bose-condensed systems in general,
covering areas such as
excitons in semiconductors,
magnon decay  \cite{Harris71,Bayrakci06,Chernyshev06},
non-linear effects in spintronics \cite{Kurebayashi11,Schultheiss09},
and strongly interacting ultracold Bose gases \cite{Papp08}.

We thank Vladimir Mineev for useful discussions,
Klaus Habicht for discussions and help in initial measurements,
and Ken Andersen and Mechthild Enderle for producing Fig.\ \ref{FigDisp}.
Invaluable technical support was obtained from Kathrin Buchner,
J\"urgen Peters, and Heinrich Kolb.
This research project has been supported by the European Commission
under the 7th Framework Programme through the `Research Infrastructures' action
of the `Capacities' Programme, Contract No: CP-CSA\_INFRA-2008-1.1.1
Number 226507-NMI3.
We acknowledge partial financial support by  ANR, contract ANR
2010-INTB-403 HIGHQ-FERMIONS (B.F.).
This work was supported, in part,  by the DOE under grant
DE-FG02-04ER46174 (A.L.C.).

\end{document}